\documentstyle[prl,aps,floats,psfig]{revtex}

\begin{document}
\twocolumn[
\hsize\textwidth\columnwidth\hsize\csname@twocolumnfalse\endcsname

\draft

\title{Spectral fluctuations of the atomic vibrations in glasses: 
random matrix theory and beyond}

\author{Jaroslav Fabian}

\address{Max-Planck Institute for the Physics of Complex Systems, 
N\"{o}thnitzer Str. 38, D-01187 Dresden, Germany}

\maketitle

\begin{abstract}
It is demonstrated on a realistic model of amorphous alloy 
Si$_{0.9}$Ge$_{0.1}$ with 1000 atoms, that short-range spectral fluctuations 
of propagons and diffusons are universal and in agreement with random matrix 
theory. The universality ceases at distances greater
than Thouless number $N_T$, where propagons obey the Altshuler-Shklovskii 
power law, $\Sigma_2\sim N^{3/2}$ for the variance 
$\Sigma_2$ of the number of levels $N$, while  a new power law, 
$\Sigma_2\sim N$, is observed for diffusons.
\end{abstract}
]
%\narrowtext
%\tighten
%\newpage
%\vspace*{1cm}

Complexity of the atomic structure of 
glasses results in a wonderful variety of vibrational
eigenmodes\cite{courtens01,fabian96,fabian97b,allen99}. 
Topological defects and the lack of 
a long-range order lead to mode localization, 
inhibition of ballistic motion, or resonance; only a small
portion of the modes resemble phonons--vibrations in periodic
crystals. It is a difficult computational
task to build a respectable-size atomic model of a glass
and obtain the vibrational eigenstructure by diagonalizing huge 
dynamical matrices; no analytical treatment is available. 
By applying the methods of random matrix theory\cite{guhr98}, this Letter
reports on certain statistical analytical rules, universal (parameter 
free) and not, shared by similar modes. In particular, it is found that
short-range spectral correlations of diffusons 
(non-propagating extended modes) 
and propagons (phonon-like modes) are universal and agree with random 
matrix theory, similarly to what have been observed in many other
complex physical systems\cite{guhr98}. However, correlations at
distances greater than Thouless number $N_T$ cannot be described
by random matrix theory. Spectral fluctuations become 
idiosyncratic: the statistics acquires a parameter ($N_T$) and the 
functional dependence is different for different mode classes. 
Propagons follow the Altshuler-Shklovskii 
``3/2'' power law \cite{altshuler86} for the fluctuations 
$\Sigma_2$ of the number of 
levels, while diffusons have the fluctuations growing 
linearly with increasing number of levels. This is likely a
general feature of strongly scattered (diffuson-like) waves in random media.

\begin{figure}
\centerline{\psfig{file=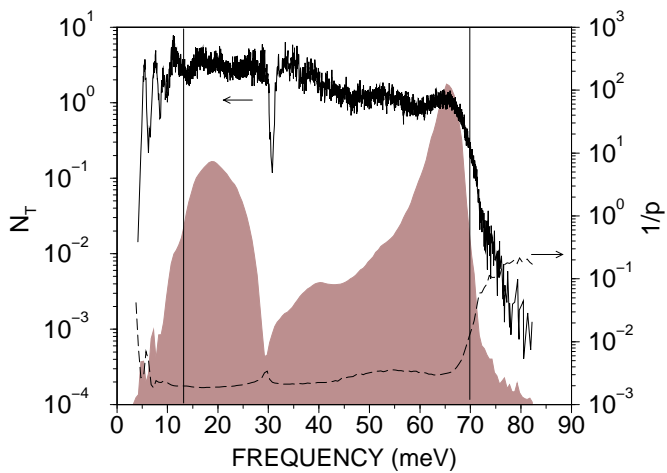,width=1\linewidth}}
\caption{Calculated vibrational density of states (shaded)
for the 1000-atom model of amorphous alloy Si$_{0.9}$Ge$_{0.1}$. 
The vertical lines are the Ioffe-Regel limit at 13 meV and 
the mobility edge at 70 meV,
dividing the spectrum into propagons (below the Ioffe-Regel limit),
diffusons (between the Ioffe-Regel limit and the mobility edge), and locons
(above the mobility edge). Also plotted is the inverse participation
ratio $1/p$ (dashed) and the Thouless number $N_T$ (solid) of the pure 
amorphous silicon model.
}\label{fig:dos}
\end{figure}

The following picture\cite{courtens01,fabian96,fabian97b,allen99} of 
vibrational modes in 
glasses has emerged from numerical studies of realistic 
models of vibrational disorder
\cite{fabian96,fabian97b,allen99,allen89,fabian97c,feldman98,biswas88}.
Figure \ref{fig:dos}
shows the calculated vibrational density of states for a model of
a glass used in this paper (amorphous Si$_{0.9}$Ge$_{0.1}$), with three 
specified spectral
regions. At the lowest frequencies vibrations are
acoustic-phonon-like propagons, weakly scattered sound waves
with well defined momentum and polarization.
At low frequencies one also finds resonance modes which have
unusually large amplitude at certain inhomogeneous regions,
while resembling propagons elsewhere. As the frequency increases, 
propagon momenta become less and less certain  until the frequency reaches
the so called Ioffe-Regel limit where the mean free path becomes
comparable to the wavelength and the wave vector concept is no longer
valid. The emerging modes are
diffusons--extended modes not able to propagate ballistically,
rather spreading out in a special diffusive fashion. Diffusons, forming
the majority of the spectrum, are the most natural modes for glasses. Atoms 
in glasses display a
short-range order by globally preserving local quantities
like interatomic distances or
coordination numbers. But at large distances the order is lost.
Diffusons have the same property: displacements of neighboring 
atoms are strongly correlated,
while those between distant atoms are uncorrelated and the global
displacement pattern, like the structure of the glass, appears
random. Finally, the highest-frequency modes are localized--locons.
The purpose of this letter is to compare spectral
properties of propagons and diffusons with random matrix theory and
theories of wave-like modes in disordered systems, to sort out 
universal and nonuniversal features.

Random matrix theory was originally devised to deal with the
spectra of complicated nuclei, but by now it has been successfully
applied in many areas of physics \cite{guhr98}.
The main assumption is
that certain spectral properties of complex (as opposed to simple)
Hamiltonians are universal, not dependent on the concrete
realization of complexity; these properties can be therefore
calculated, in many cases analytically,  as averages over an
ensemble of (similarly) complex Hamiltonians. I focus here on the
Gaussian orthogonal ensembles (GOE) which are sets of real
symmetric matrices with elements randomly selected from a Gaussian
probability distribution. Dynamical matrices which determine the
frequencies of atomic vibrations are also real symmetric so, if
their structure is sufficiently complicated, their properties
should be similar to the properties of the GOE. For the latter
theory predicts that neighboring levels (frequencies) repel each
other: there is a zero probability of finding two equal levels. A
convenient measure of level repulsion is the level-spacing
distribution (LSD) $\rho$ which, for the GOE ensembles, is very
accurately expressed by the Wigner surmise\cite{guhr98}
\begin{eqnarray}\label{eq:wigner}
\rho_W(s)=\frac{\pi}{2}s\exp(-\frac{\pi}{4}s^2),
\end{eqnarray}
with $s$ denoting the level spacing. Random
matrix theory has also predictions about correlations between more
distant levels. Such correlations can be quantified in several
ways. Here I use the $\Sigma_2$ statistics which
measures, given an interval of size $N$, the variance of the 
number of levels in that interval. For GOE at large $N$
\cite{guhr98}:
\begin{eqnarray}\label{eq:sigma2}
\Sigma_{2,\rm GOE}(N)\approx\frac{2}{\pi^2}\left[\ln(2\pi
N)+\gamma+1-\frac{\pi^2}{8}\right],
\end{eqnarray}
where $\gamma\approx0.5772$ is Euler's constant. The logarithmic
dependence expresses the rigidity of a GOE spectrum.

I evaluate LSD and $\Sigma_2$ for the vibrational spectrum of 
a model of amorphous Si$_{0.9}$Ge$_{0.1}$ alloy whose atomic
coordinates and interatomic forces are taken from the realistic 
model of amorphous silicon\cite{allen89,broughton87} generated 
by randomizing a crystalline silicon
structure (Wooten-Winer-Weaire recipe\cite{wooten85}) 
and then relaxing the atoms to a local minimum of the
Stillinger-Weber interatomic potential \cite{stillinger85}. 
The model structure and physical properties agree
very well with experiment \cite{allen99,allen89,fabian97c}. 
Here I use a model with 1000 atoms arranged in a cube of side 27 \AA.
The Si$_{0.9}$Ge$_{0.1}$ alloy is obtained by substituting germanium atomic
masses for the masses of randomly picked 10\% silicon atoms. The 
mass disorder is small to qualitatively change the vibrational 
properties of the original amorphous silicon model (similar
amorphous silicon-germanium alloys were studied in 
Refs.\cite{allen89,bouchard88}), but large enough 
to generate a useful ensamble. Below I use $N_r=3500$ different mass-disorder
realizations of the alloy. The model has 3000 vibrational frequencies 
whose spectrum is in Fig.\ref{fig:dos} (averaged over the ensamble),
along with the alloy inverse participation ratio $1/p$ (averaged over 15
samples), allowing an identification of the mobility edge at about 70 meV. 
The Ioffe-Regel limit is also indicated at 13 meV (taken from the studies of the
pure silicon models\cite{allen89}; the actual transition is probably 
quite broad, perhaps
plus and minus 2 meV). For the statistics, I take for propagons all the
modes, about 200, from 5 to 13 meV (this includes a small number, less 
than 5\%, of resonance modes which behave mostly as propagons anayway) and
for diffusons all the modes, about 1000, from 35 to 65 meV (which is 
enough away from the locons and from the anomalous quasi-localized region 
around 30 meV at the DOS dip and 1/p peak).

\begin{figure}
\centerline{\psfig{file=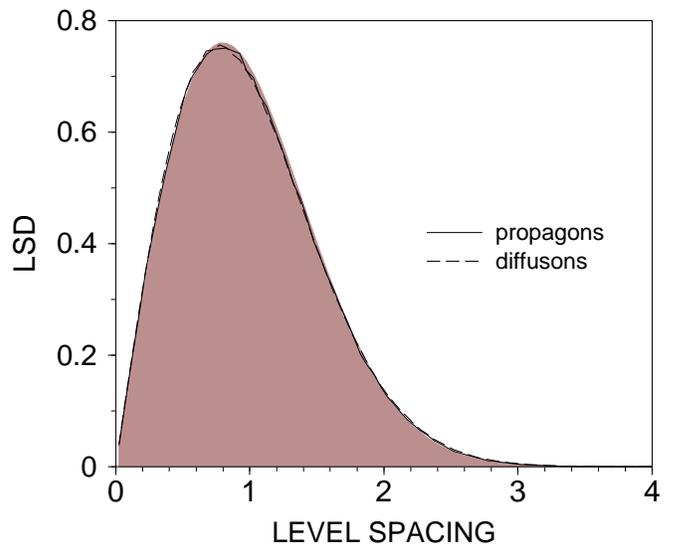,width=1.0\linewidth}}
\caption{Calculated level-spacing distribution (LSD) for propagons
and diffusons, and the GOE result (shaded), Eq.\ref{eq:wigner}.} 
\label{fig:lsd}
\end{figure} 

To calculate spectral averages that can be compared with the
predictions of random matrix theory, the vibrational spectrum has
to be ``unfolded,'' that is, locally rescaled to compensate for
the spectral variations of the local average level spacing. A
standard unfolding procedure \cite{guhr98} involves finding 
the average cumulative DOS (a staircase function) of the
spectral region of interest. The unfolded spectrum is then the set
of values of the average function at the original frequencies. 
There are two ways of finding the average: by ensamble averaging or
by analyzing a smooth fit to a single spectrum. The latter method
works fine if an analytical behavior of the cumulative DOS
is known. Without an analytical guidance, however, the procedure is more 
art than science\cite{guhr99}. This is why I use amorphous alloy 
Si$_{0.9}$Ge$_{0.1}$ (rather than just pure amorphous silicon) for spectral
statistics: first, there is no analytical formula available for DOS of 
amorphous silicon and, second, mass disorder is much easier to implement
than structural disorder. 
With a sufficiently large ensamble
at hand, the averages are straightforward to evaluate. If $\rho^i_c(\omega)$
is the cumulative DOS of the $i$-th sample at frequency $\omega$, 
the average is $\rho_c(\omega)=(1/N_r)\sum_i \rho^i(\omega)$. After
obtaining $N_r$ unfolded spectra, LSD and $\Sigma_2$ for
a certain frequency range are calculated by averaging over both the
spectral region and the ensamble\cite{levels}. 

\begin{figure}
\centerline{\psfig{file=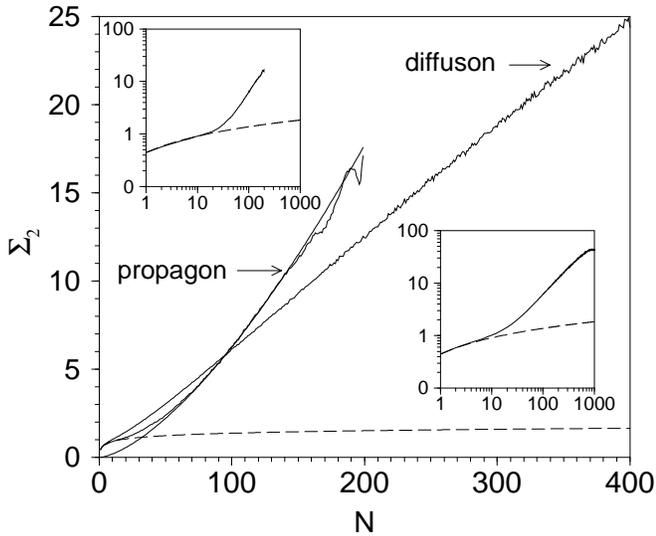,width=1.0\linewidth}}
\caption{Calculated $\Sigma_2$ for propagons and diffusons.
The dashed line is the GOE result and the thin line accompanying
the propagon curve is  $\Sigma_2=0.0076$ $ \times (N/1.1)^{3/2}$, 
the Altshuler-Shklovskii power law, Eq. \ref{eq:as}, with effective 
Thouless number 1.1. The insets are the log-log plots for
propagons (up) and diffusons (down).}
\label{fig:sigma2}
\end{figure}

The calculation of LSD for both propagons and diffusons is in Fig.
\ref{fig:lsd}. The agreement with the Wigner surmise is excellent,
proving the level repulsion between neighboring modes and, once again, 
the robustness of the LSD predictions. This connection between
vibrations in glasses and GOE matrices was already reported in Ref.
\cite{fabian97b} (and in Ref. \cite{schirmacher98} for other 
disordered-lattice systems). It turns out, however, that random
matrix theory is not obeyed by the correlations
between more distant than neighboring (as in LSD) levels.

Consider first propagons. The calculated $\Sigma_2$ is in Fig. 
\ref{fig:sigma2}. Deviations from random matrix theory, 
starting already at small $N$, grow as a power law rather than 
a logarithm. The power law, which extends from $N\approx 40$ to 
200 is identified as $\Sigma_2\sim N^{3/2}$ (best seen in the inset), 
and is precisely the one predicted
by Altshuler and Shklovskii\cite{altshuler86}, originally for weakly
scattered electrons on a 3-dimensional disordered lattice: 
\begin{eqnarray}\label{eq:as}
\Sigma_3(N)\approx 0.0076\times (N/N_T)^{3/2};
\end{eqnarray}   
the exact proportionality constant is $\sqrt{2}/6\pi^3 \approx 0.0076$.
The scaling parameter $N_T$ is the Thouless number, $N_T = \hbar D/\delta L^2$, 
the uncertainty in the energy (in the units of the average local 
level spacing $\delta$) of a wave-packet-like state diffusing with 
diffusivity $D$ throughout a sample of size $L$. Equation \ref{eq:as} 
is valid for $N \gg N_T$ \cite{altshuler86} and propagons obey it with 
$N_T\approx 1.1$ (see Fig. \ref{fig:sigma2}).
Figure \ref{fig:dos} plots $N_T$ as a function of vibrational frequency for
the pure silicon 1000-atom model, calculated from the model diffusivity 
$D$\cite{allen89}. The result for the alloy would look identical, perhaps 
being smaller by a few percent ($D$ for a similar alloy with 25\% of 
substituted Ge atoms differs only slightly from $D$ of the pure silicon 
sample \cite{allen89}). We see that $N_T$ 
is frequency dependent--it varies from below 0.1 to above 4 in the propagon
region. Considering that Eq. \ref{eq:as} comes with some effective 
$N_T$, biased towards lower values, the result for the effective 
$N_T\approx 1.1$ is reasonable (the average of $1/N_T^{3/2}$ over 
the propagon frequencies gives about 1.7; for the alloy it will be 
slightly less), confirming the theory of Ref.\cite{altshuler86}.

\begin{figure}
\centerline{\psfig{file=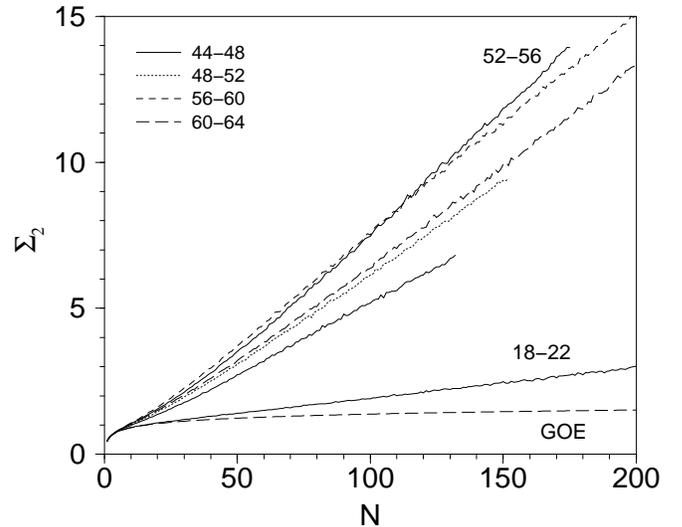,width=1\linewidth,angle=0}}
\caption{Spectral statistics $\Sigma_2$ of selected diffuson regions.
The intervals in the legend and with the two of the curves are the spectral
ranges in meV (see DOS in Fig.\ref{fig:dos}). The GOE curve is also shown. 
The Thouless numbers of
the regions are: 3.2 (18-22), 1.2 (44-48), 1.3 (48-52), 1.2 (52-56),
0.96 (56-60), and 1 (60-64). The corresponding slopes of the linear
increase are: 0.011, 0.049, 0.063, 0.085, 0.074, and 0.068.
For example, diffusons from 44 to 48 meV  have  
$N_T\approx 1.2$, and at large $N$ their $\Sigma_2\approx 0.049\times N$.
}
\label{fig:sigma_all}
\end{figure}

Diffusons show nonuniversal behavior too. In Fig. \ref{fig:sigma2},
their $\Sigma_2$ grows linearly with increasing $N$ for about a decade,
from 30 to almost 400 (the growth slows down for greater $N$, most 
likely due to finite size effects). Like propagons, diffusons
become less spectrally correlated when separated by more than a few 
other modes (of order 1; the deviation at small $N$ is not visible 
because of the graph scale).  But unlike propagons, the power law is linear, 
$\Sigma_2\sim N$. This different dependence is not surprising, as 
diffusons
are {\it not} weakly scattered modes envisioned in the derivation of Eq.
\ref{eq:as}. Diffusons do diffuse with time \cite{allen98},
but in a different way than do propagons or electrons (in weakly 
disordered systems). Equation \ref{eq:as} was derived by perturbation
theory which assumes electron states to be plane waves between
scattering events. In other words, the mean free path of electrons
is much greater than their wavelength: the electrons are below the
Ioffe-Regel limit and their diffusion resembles a random walk (like
propagons).  Diffusons, on the other hand, diffuse more like a 
free-particle quantum mechanical wave packet (this type  
of diffusion was branded ``intrinsic'' \cite{allen89}); diffusons are 
not perturbative states.

To explore possible sources of differences 
between $\Sigma_2$ for propagons and diffusons, it is instructive 
to repeat the qualitative argument \cite{altshuler86} 
which leads to Eq. \ref{eq:as}. A given state can mix with neighboring $N
\gg N_T$ states if it diffuses a time $\tau_N\approx \hbar/N\delta$, or over a
region with size $L_N=(D\tau_N)^{1/2}=(N_T/N)^{1/2}L$. The
$N$ states in a volume $(L_N)^3$ will be correlated as
predicted by random matrix theory, with the uncertainty in the
number of levels of order unity (a rigid spectrum). But states
in different regions (all of size $L_N$) of the sample
will be uncorrelated, so the uncertainty (measured by $\Sigma_2$)
in the number of states in the
whole sample will be proportional to the number of the independent
regions/spectra. This number is $(L/L_N)^3 = (N/N_T)^{1.5}$ which
is Eq. \ref{eq:as}. The argument can be made (naively) more general by 
considering an anomalous diffusive transport 
$L_N\sim \tau^{\alpha}_N$ ($\alpha=1/2$ in the usual case) 
and a fractal filling of volume 
$L^{\beta}$ ($\beta=3$ for uniform filling). Then 
$\Sigma_2\sim N^{\alpha\beta}$, and for diffusons one gets
$\alpha\beta=1$. If the space filling is ordinary ($\beta=3$),
the diffusive transport is not ($\alpha=1/3$). If the transport is
ordinary ($\alpha=1/2$), the spreading is fractal ($\beta=2$). It is
not clear wheather diffusons behave one way or the other (or both),
or whether the above reasoning, which is a pure speculation,
makes sense. 

If there is a scaling of the type $\Sigma_2\sim (N/N_T)$, the best
way to see it is to look at spectral statistics of small diffuson regions
of almost constant $N_T$; the regions must contain enough levels
to see the systematic deviations from GOE. Figure \ref{fig:sigma_all}
plots $\Sigma_2$ of six such regions, all displaying the
linear law $\Sigma_2\sim N$, all with different proportionality
coefficients (note that 
the law holds also for the 18-22 region not considered in the
calculation of Fig. \ref{fig:sigma2}). From the values of the 
coefficients and the regions'
average $N_T$'s given in the caption, one cannot see any obvious 
relationship between the coefficients and $N_T$. There is certainly
no support from the available data of a scaling
of the type $\Sigma_2\sim (N/N_T)$: by plotting $\Sigma_2$
versus $N/N_T$ (not shown) the curves do not collapse onto or close to 
a single
line. While there definitely is a dependence of $\Sigma_2$ on 
$N_T$ (explicit or implicit), the present spectrum size does not 
allow me to decipher it and considering larger models is presently 
unfeasible.

I thank P. B. Allen, J. Feldman, and A.
M. Halasz for stimulating discussions.

\end{document}